\documentclass[aps,prd,onecolumn,superscriptaddress,showpacs,floatfix,10pt]{revtex4}
\usepackage{graphicx}% Include figure files
\usepackage{amssymb,amsmath}
\usepackage{subfigure}
\usepackage{epsfig}

\begin{document}

\title{Quantum dynamics of massive particles in a non-commutative two-sheeted
space-time}

\author{Fabrice Petit}
\email{f.petit@bcrc.be} \affiliation{Belgian Ceramic Research
Centre,\\4 avenue du gouverneur Cornez, B-7000 Mons, Belgium}

\author{Micha\"{e}l Sarrazin}
\email{michael.sarrazin@fundp.ac.be} \affiliation{Laboratoire de
Physique du Solide, \\Facult\'es Universitaires Notre-Dame de la
Paix, \\61 rue de Bruxelles, B-5000 Namur, Belgium}

\date{\today}

\begin{abstract}
We study a formal extension of the Dirac equation in the framework
of a non-commutative two-sheeted space-time. It is shown that this
approach naturally extends the classical Dirac theory by doubling
the number of fermionic states, which can then be identified as
matter and hidden-matter states. Our model exhibit several
interesting features that could have observational consequences.
Among them, we predict a small electromagnetic coupling between
matter and hidden matter universes which should lead to
matter/hidden matter oscillations in presence of intense
electromagnetic vector potentials.
\end{abstract}

%%%%%%%%%%%%%%%%%%%%%%%%%%%%%%%%%%%%%%%%%%%%%%%%%%

\maketitle
%\newpage
\section{Introduction}

The concept of hidden matter traces back to 1956 when Lee {\&}
Yang first noticed that the parity violation problem involved in
the weak interactions could be solved by enlarging the particle
content to include a mirror sector [1]. The underlying idea of a
mirror sector is to duplicate the standard model and allow
opposite symmetry breaking in the two sectors. Thus, for each
left-handed particle there would be a right-handed mirror partner
to restore parity.

This idea was further extended by a number of authors over the
years and there is now a huge collection of papers devoted to this
topic. Usually, it is assumed that usual matter and matter from
the hidden world can not interact through ordinary interactions
except gravitation. As a consequence, hidden matter made of hidden
atoms could exist with exactly the same internal properties as
ordinary matter but would be completely undetectable for us
through electromagnetic means. In recent years however, Foot {\&}
Volkas have suggested to extend the original idea to allow a
possible coupling of matter and mirror matter at the quantum
level. This coupling involves some specific kind of interactions
including for instance photon-mirror photon kinetic mixing [2] and
also neutrino-mirror neutrino mass mixing [3]. These authors
conclude that even if those interactions are tiny, the
experimental consequences could be dramatic. Several possible
astrophysical, cosmological and physical implications of mirror
matter are extensively reviewed in [2].

Many other theoretical approaches use the phenomenological power
of a hidden sector. One of the most important concerns superstring
theories with $E8\times E8$ symmetry group [5]. This approach
assumes that particles are associated with the endpoints of open
strings which are attached to D-branes. Ordinary and hidden-sector
particles live then on different branes embedded in the bulk of a
higher-dimensional space.

More recently, A. Connes proposed a two-sheeted space-time using
non-commutative geometry (NCG) [6,7]. In his work, left and
right-handed fermions are assumed to live on the two different
sheets, which are coupled by a scalar field representing the Higgs
field. Thus, in the low energy limit, this theory predicts the
existence of two copies of space-time associated with a double
Hilbert space. The cornerstone of this approach is to restrict the
five dimensional space to a finite number of discrete points,
generally two [8-12]. In several aspects, the two-sheeted
space-time represents a discretized version of Kaluza-Klein theory
in which, the fifth compact circular dimension is replaced by
discrete points. This theory also presents some specific
advantages like for instance a possible explanation of the huge
difference between the electroweak and the Planck scales.

In the present paper, the NCG developed by A.Connes will be used
to extend further the idea of a hidden sector embedded in a 5D
bulk. Our study focuses on the dynamics of a massive particle in a
two-sheeted space-time using relevant extensions of the Dirac and
Pauli equations. The results of this model differ from previous
works of literature essentially by the way of the particle mass is
introduced into the model. It is shown that this approach leads to
several interesting phenomena that could have strong observational
consequences. The most noticeable ones concern two-sheeted
oscillations of massive fermions in presence of an electromagnetic
vector potential and a possible increase of the electric charge
with the particle velocity.

The paper is organized as follows. In Sec. 2, we will first
develop the minimal mathematical knowledge required to introduce
the NC two-sheeted space-time. Then, we shall propose a formal
extension of the Dirac equation and we will show that it allows
for exact diagonalization in flat space-time. In Sec. 3, the
non-relativistic limit of the NC Dirac equation in presence of a
two-sheeted gauge field will be derived. It will be shown that it
leads to a system of coupled Pauli equations relative to both
sheets. Finally, we will determine the effect on the particle
dynamics of 1) a constant magnetic vector potential solely or 2)
coupled with a magnetic field. The last section closes by
discussing some physical implications of the model.

\section{Z2 Non-commutative Dirac equation in flat space-time --
Mathematical framework}

Let us introduce a Bi-Euclidean space $X$ defined as the product
of a 4 continuous manifold by a discrete two points space, i.e.
$X=M\times Z_2$. Any smooth function in this spacetime belong to
the algebra $A=C^\infty (M)\oplus C^\infty (M)$ and can be
adequately represented by a 2x2 diagonal matrix $F$ such that :
\begin{equation}
F =\left( {{
\begin{array}{cc}
{f_1 } & 0 \\
0 & {f_2 }
\end{array}
}}\right) \label{eq1}
\end{equation}
The expression of the exterior derivative $D=d+Q$ where d acts on
M and Q on the $Z_2$ internal variable has been given by A. Connes
[6] : $D:(f_1,f_2)\rightarrow (df1,df2,\delta .(f_2-f_1),\delta
.(f_1-f_2))$. Viet has proposed a representation of D acting as a
derivative operator and fulfilling the above requirements [9]
Following this author we set :
\begin{equation}
D_\mu =\left( {{
\begin{array}{cc}
{\partial _\mu } & 0 \\
0 & {\partial _\mu }
\end{array}
}}\right) ,\text{ }\mu =0,1,2,3\text{ and\ }D_5=\left( {{
\begin{array}{cc}
0 & \delta  \\
{-\delta } & 0
\end{array}
}}\right)   \label{eq1}
\end{equation}
Throughout this paper, we work in units with $\not{h}=c=1$.

The parameter $\delta $ has the dimension of mass in order to give
the fifth component of space-time the same dimension as the other
components. Its acts as a finite difference operator along the
discrete dimension and corresponds formally to the distance
between the two sheets. Using (2), one can build the Dirac
operator defined as
\begin{equation}
\not{D}=\Gamma ^ND_N=\Gamma ^\mu D_\mu +\Gamma ^5D_5  \label{eq2}
\end{equation}
By considering the following extension of the gamma matrices (we
are working in the Hilbert space of spinors, see [7])
\begin{equation}
\Gamma ^\mu =\left( {{
\begin{array}{cc}
{\gamma ^\mu } & 0 \\
0 & {\gamma ^\mu }
\end{array}
}}\right) \text{\ and\ }\Gamma ^5=\left( {{
\begin{array}{cc}
{\gamma ^5} & 0 \\
0 & {-\gamma ^5}
\end{array}
}}\right)   \label{eq3}
\end{equation}
it can be easily shown that the Dirac operator given by eq.3 has
the following self adjoint realization :
\begin{equation}
\not{D}=\left( {{
\begin{array}{cc}
{\not{D}_{+}} & {\delta \gamma ^5} \\
{\delta \gamma ^5} & {\not{D}_{-}}
\end{array}
}}\right) =\left( {{
\begin{array}{cc}
{\gamma ^\mu \partial _\mu } & {\delta \gamma ^5} \\
{\delta \gamma ^5} & {\gamma ^\mu \partial _\mu }
\end{array}
}}\right)   \label{eq4}
\end{equation}
Usually in Z2 non commutative geometry, one considers that the off
diagonal term proportional to $\delta$ (which is a matrix in the
most generalized case) is related to the particle mass through the
Higgs field. In this work we make a different choice and consider
that $\delta$ is constant and take the same value for every
particles. To take into account different particles, it is thus
necessary to introduce a mass term as in the classical approach of
Dirac's equation, e.g:
\begin{equation}
M=m \left( {{
\begin{array}{cc}
{I_4} & 0 \\
0 & {I_4}
\end{array}
}}\right)   \label{eq5}
\end{equation}
whereas we keep the scalar field $\delta $ constant.

By analogy with the classical approach, one can then construct a
2-sheeted Dirac equation such that
\begin{equation}
D_{irac} \Psi =\left( {i\not{D}-M}\right) \Psi =0 \label{eq6}
\end{equation}
with $\Psi =\left( {{
\begin{array}{c}
{\psi _{+}} \\
{\psi _{-}}
\end{array}
}}\right) $ the two sheeted wave function. In this notation, the
indices ``$+$'' and ``$-$'' are purely conventional and simply
allow to discriminate the two sheets embedded in the 5D bulk.

It can be easily shown that the equation (7) can be derived from
the Lagrangian \textbf{\textit{L}} such that
\begin{equation}
\textbf{\textit{L}}=\bar{\Psi} \left( {i\not{D}-M}\right) \Psi
\label{eq7}
\end{equation}
Where $\bar{\Psi}=\left( {\bar{\psi}_{+},\bar{\psi}_{-}}\right) $
is the two
sheeted spinor adjoint to $\Psi $ and with $\bar{\psi}_{+}$ and $\bar{\psi}%
_{-}$ the spinors conjugated respectively to $\psi _{+}$ and $\psi
_{-}$.

From (7), one gets~
\begin{equation}
\left( {i\not{D}-M}\right) \Psi =\left( {i\Gamma ^ND_N-M}\right)
\Psi =\left( {{
\begin{array}{cc}
{i\gamma ^\mu \partial _\mu -m} & {i\delta \gamma ^5} \\
{i\delta \gamma ^5} & {i\gamma ^\mu \partial _\mu -m}
\end{array}
}}\right) \left( {{
\begin{array}{c}
{\psi _{+}}  \\
{\psi _{-}}
\end{array}
}}\right)   \label{eq8}
\end{equation}
Thus the equation of motion for the two-component field $\left( {{
\begin{array}{c}
{\psi _{+}} \\
{\psi _{-}}
\end{array}
}}\right) $ becomes
\begin{equation}
i\gamma ^\mu \partial _\mu \psi _{+}-m\psi _{+}+i\delta \gamma ^5
\psi _{-}=0  \label{eq9}
\end{equation}
\begin{equation}
i\gamma ^\mu \partial _\mu \psi _{-}-m\psi _{-}+i\delta \gamma ^5
\psi _{+}=0  \label{eq10}
\end{equation}
The coupling between the two sheets arises from the presence of
the $\delta $ term. It disappears completely for $\delta =0$,
which corresponds to infinitely, separated sheets. Note that the
system derived here is similar to the one discussed in [13] to
explain the flavor oscillation of neutrinos. We will return to
this in the last section. It is worth noticing that if the mass m
equals to zero, then the equations 10 and 11 turn out to be the
standard equations of the Z2 non-commutative spacetime [6]. In
that case, the indices $+$ and $-$ can be substituted by the
indices $L$ and $R$ which refer to actual left and right parity
states.

\section{Free field solution of the Z2-Dirac equation}

To solve the system (9), we introduce the auxiliary field $\Phi $
such that
\begin{equation}
\Psi =\left\{ {i\Gamma ^ND_N+M}\right\} \Phi   \label{eq11}
\end{equation}
It is then straightforward to show that $\Phi $ satisfies
\begin{equation}
\left\{ {\square+m^2+\delta ^2}\right\} \Phi =0 \label{eq12}
\end{equation}
Let us make the following plane wave solution ansatz: $\Phi =\Phi
_0\exp \left[ {-i\varepsilon \left( {Et-\vec{p}\cdot
\vec{x}}\right) } \right] $ with $\varepsilon =\pm 1$ to take into
account the positive and negative energy solutions.

Then eq.13 gives
\begin{equation}  \label{eq13}
\left[ {-E^2+p^2+m^2+\delta ^2} \right]\Phi =0
\end{equation}
such that the energy eigenvalues are $E=\sqrt {p^2+m^2+\delta ^2} $with $%
\varepsilon =\pm 1$. Note that the distance between the two sheets
appears here as a simple correction to the particle rest mass. So,
even in the case where $m=0$, the mass of the fermion can never be
completely equals to zero. Perhaps, the smallness of neutrino
masses could be explained that way.

For a momentum
\begin{equation}
\vec{p}=p\left( {\sin \theta \cos \phi ,\sin \theta \sin \phi
,\cos \theta } \right)   \label{eq14}
\end{equation}
we now define two component eigenstates of the matrix
$\vec{\sigma}\cdot \vec{p}$ for later convenience
\begin{equation}
\chi _{1/2}\left( {\vec{p}}\right) =\left( {{
\begin{array}{c}
{\cos \left( {\theta /2}\right) } \\
{\sin \left( {\theta /2}\right) \exp \left[ {i\phi }\right] }
\end{array}
}}\right) \text{\ and\ }\chi _{-1/2}\left( {\vec{p}}\right)
=\left( {{
\begin{array}{c}
{-\sin \left( {\theta /2}\right) \exp \left[ {-i\phi }\right] } \\
{\cos \left( {\theta /2}\right) }
\end{array}
}}\right)   \label{eq15}
\end{equation}
For the positive energy, we then look for solutions of the form
\begin{equation}
\Phi _\lambda =\left[ {{
\begin{array}{c}
{\chi _\lambda } \\
0 \\
0 \\
0
\end{array}
}}\right] \text{\ or\ }\Phi _\lambda =\left[ {{
\begin{array}{c}
0 \\
0 \\
{\chi _\lambda } \\
0
\end{array}
}}\right]   \label{eq16}
\end{equation}
whereas for the negative energy, we consider
\begin{equation}
\Phi _\lambda =\left[ {{
\begin{array}{c}
0 \\
{\chi _\lambda } \\
0 \\
0
\end{array}
}}\right] \text{\ or\ }\Phi _\lambda =\left[ {{
\begin{array}{c}
0 \\
0 \\
0 \\
{\chi _\lambda }
\end{array}
}}\right]   \label{eq17}
\end{equation}
with $\lambda =\pm 1/2$.

Using the following representation of the Dirac matrices
\begin{equation}
\gamma ^0=\left( {{
\begin{array}{cc}
{I_2} & 0 \\
0 & {-I_2}
\end{array}
}}\right) ,\text{ }\gamma ^i=\left( {{
\begin{array}{cc}
0 & {\sigma _i} \\
{-\sigma _i} & 0
\end{array}
}}\right) ,\text{ }\gamma ^5=\left( {{
\begin{array}{cc}
0 & {I_2} \\
{I_2} & 0
\end{array}
}}\right)   \label{eq18}
\end{equation}
with $\left( {\gamma ^5}\right) ^2=1$ and $\left\{ {\gamma
^5,\gamma ^\mu } \right\} =0$, it can be shown that eq. 12 leads
to
\begin{equation}  \label{eq19}
\Psi =\left[ {{
\begin{array}{cccc}
{E+m} & {-\vec {\sigma }\cdot \vec {p}} & 0 & {i\delta } \\
{\vec {\sigma }\cdot \vec {p}} & {-E+m} & {i\delta } & 0 \\
0 & {i\delta } & {E+m} & {-\vec {\sigma }\cdot \vec {p}} \\
{i\delta } & 0 & {\vec {\sigma }\cdot \vec {p}} & {-E+m}
\end{array}
}} \right]\Phi
\end{equation}
Using eq. 17 and 18, the solutions for $\Psi $ can be easily
derived. After normalization, one gets

for the positive energy ($\varepsilon =+1$)
\begin{equation}
U_\lambda =\frac 1{\sqrt{2E \left( {E+m}\right) }} \left[ {{
\begin{array}{c}
{\left( {E+m}\right) \chi _\lambda } \\
{\vec{\sigma}\cdot \vec{p}\chi _\lambda } \\
0 \\
{i\delta \chi _\lambda }
\end{array}
}}\right] ,\text{ }\tilde{U}_\lambda =\frac 1{\sqrt{2E \left(
{E+m} \right) }} \left[ {{
\begin{array}{c}
0 \\
{i\delta \chi _\lambda } \\
{\left( {E+m}\right) \chi _\lambda } \\
{\vec{\sigma}\cdot \vec{p}\chi _\lambda }
\end{array}
}}\right]   \label{eq20}
\end{equation}
and for the negative energy ($\varepsilon =-1$)
\begin{equation}
V_\lambda =\frac 1{\sqrt{2E \left( {E+m}\right) }} \left[ {{
\begin{array}{c}
{\vec{\sigma}\cdot \vec{p}\chi _\lambda } \\
{\left( {E+m}\right) \chi _\lambda } \\
{i\delta \chi _\lambda } \\
0
\end{array}
}}\right] ,\text{ }\tilde{V}_\lambda =\frac 1{\sqrt{2E \left(
{E+m} \right) }} \left[ {{
\begin{array}{c}
{i\delta \chi _\lambda } \\
0 \\
{\vec{\sigma}\cdot \vec{p}\chi _\lambda } \\
{\left( {E+m}\right) \chi _\lambda }
\end{array}
}}\right]   \label{eq21}
\end{equation}
It is instructive to compare those solutions with the usual ones
given by the standard Dirac equation, i.e. for the positive energy
\begin{equation}
u_\lambda =\frac 1{\sqrt{2E \left( {E+m}\right) }}\left[ {{
\begin{array}{c}
{\left( {E+m}\right) \chi _\lambda } \\
{\vec{\sigma}\cdot \vec{p}\chi _\lambda }
\end{array}
}}\right]   \label{eq22}
\end{equation}
and for the negative energy
\begin{equation}
v_\lambda =\frac 1{\sqrt{2E \left( {E+m}\right) }}\left[ {{
\begin{array}{c}
{\vec{\sigma}\cdot \vec{p}\chi _\lambda } \\
{\left( {E+m}\right) \chi _\lambda }
\end{array}
}}\right]   \label{eq23}
\end{equation}
One can see that the two-sheeted solutions can be identified with
the classical ones provided that $\delta \to 0$. So, for a very
small $\delta $, the difference between the standard and the two
sheeted Dirac theory is not expected to be significant. In our
approach the positive and negative energy solutions are assumed to
correspond to particle and antiparticle respectively, as in the
classical Dirac theory.

The form of solutions $\Psi $ indicates that the fermion doubling
is related to the two possible localization of the particles in
5D, i.e. in one or the other sheet. For illustrative purpose, let
us consider the case of a positive energy particle. A particle
mainly located in the first sheet can be written as a linear
combination of $U_\lambda $ solutions, i.e.
\begin{equation}
\Psi =\frac 1{\sqrt{V}}\sum\limits_\lambda {N_\lambda U_\lambda }
\exp \left[ {-i\left( {Et-\vec{p}\cdot \vec{x}}\right) }\right]
\label{eq24}
\end{equation}
with $N_\lambda $ such that $\sum\limits_\lambda {\left|
{N_\lambda }\right| ^2}=1$.

Let also $P_{+}$ and $P_{-}$ be the probability for the particle
to be in the first (noted $+$) and second (noted $-$) sheet
respectively. Then, considering the integrated value of  $\left|
{\psi _{+}}\right| ^2$ and $ \left| {\psi _{-}}\right| ^2$ (given
by the four first and four last components of $\Psi $), it can be
shown that
\begin{equation}
P_{+}=1-K  \label{eq25}
\end{equation}
\begin{equation}
P_{-}=K  \label{eq26}
\end{equation}
with $K=\delta ^2/\left[ {2E \left( {E+m}\right) }\right] $.
Provided that $\delta $ is small enough, i.e. $K<<1$, one verifies
that the particle is mainly in the sheet ``$+$''. In the same way,
a particle corresponding to
a wave function $\Psi $ written as a linear combination of $\tilde{U}%
_\lambda $ solutions is mainly located in the sheet ``$-$''.
Notice that since the ``confinement'' of the particle within the
sheet increases with the energy of the particle (decreasing $K$),
the apparent electric charge of a particle should follow the same
behavior and this could be a possible way to experimentally check
out the validity of the model. We note that if we consider, for
instance, the case of an electron in the rest of frame and if we
suppose a distance between both sheet on the order of $1$ mm, the
related value of $\delta $ leads to $K\sim 4\cdot 10^{-20}$. Even
if the distance decreases to one angstrom, then $K\sim 4\cdot
10^{-6}$. The electric charge values $q=eK$ then obtained are
close but out of the range of measurements obtained in the current
experiments [14].

So, in the two-sheeted spacetime, positive energy fermions can be
localized indifferently in one or the other sheet. A similar
consideration holds of course for a negative energy particle.

\section{Role of the electromagnetic field and the Pauli equation in the two
sheeted space-time}

The most general form for a two-sheeted gauge field is given by
\begin{equation}
A=\left( {{
\begin{array}{cc}
{A_{+}} & {\chi} \\
{\chi} & {A_{-}}
\end{array}
}}\right)   \label{eq27}
\end{equation}
A gauge field on such a generalized space-time consists of the
usual gauge field $A_{+}$, $A_{-}$ in the two Euclidean manifolds
supplemented by a scalar field $\chi $, usually identified as the
Higgs field. In this paper, we will limit ourselves to the more
restrictive gauge with $\chi =0$ to concentrate only on the effect
of the photon fields.

By construction, $A_{+}$ is coupled with the four first components
of the spinor $\Psi $ (i.e. $\psi _{+}$) whereas $A_{-}$ is
coupled with the four last components (i.e. $\psi _{-}$). The
minimal coupling of the gauge field with the Dirac fields yields
to
\begin{equation}
\left( {i\left( {\not{D}+A}\right) -M}\right)\Psi =\left( {{
\begin{array}{cc}
{i\gamma ^\mu \left( {\partial _\mu +iA_{+,\mu }}\right) -m} &
{i\delta
\gamma ^5} \\
{i\delta \gamma ^5} & {i\gamma ^\mu \left( {\partial _\mu
+iA_{-,\mu }} \right) -m}
\end{array}
}}\right) \left( {{
\begin{array}{c}
{\psi _{+}} \\
{\psi _{-}}
\end{array}
}}\right)   \label{eq28}
\end{equation}
It is instructive to note that the operator $\Gamma ^5D_5$ in $
\not{D}$ does not commute with $A$ since
\begin{equation}
\left[ {\Gamma ^5D_5,A}\right] =\left( {{
\begin{array}{cc}
0 & {2i\delta \gamma ^\mu \gamma ^5\left( {A_{+\mu }-A_{-\mu
}}\right)
} \\
{2i\delta \gamma ^\mu \gamma ^5\left( {A_{-\mu }-A_{+\mu }}\right)
} & 0
\end{array}
}}\right)   \label{eq29}
\end{equation}
This point is noteworthy as the off-diagonal terms suggests the
existence of electromagnetic coupling between matter and hidden
matter sectors. Notice that in order to avoid a breaking of the
electromagnetic gauge invariance, we are forced to consider that
the same gauge transformation applies simultaneously to both
sheets. As the value of $\delta $ is not dictated by any obvious
physical considerations, the coupling strength might be strong
enough to affect quantum phenomena. The last part of this paper
addresses this important issue.

To clarify the effect of the coupling term, it is suitable to
derive the non-relativistic limit of the Dirac equation. We start
with
\begin{equation}  \label{eq30}
i\gamma ^\mu \left( {\partial _\mu +iA_{+,\mu } } \right)\psi _+
-m\psi _+ +i\delta \gamma ^5\psi _- =0
\end{equation}
\begin{equation}  \label{eq31}
i\gamma ^\mu \left( {\partial _\mu +iA_{-,\mu } } \right)\psi _-
-m\psi _- +i\delta \gamma ^5\psi _+ =0
\end{equation}

Following the standard procedure to derive Pauli's equation from
Dirac's one, one can easily show that :
\begin{equation}
i\frac{\partial \varphi _1}{\partial t}=\left[ {\frac 1{2m}\left[ {\left( {-i%
\vec{\nabla}-\vec{A}_{+}}\right) ^2+\delta ^2}\right] +A_{+,0}-\left( {\frac{%
\vec{\sigma}\cdot \vec{B}_{+}}{2m}}\right) }\right]\varphi
_1+i\frac
\delta {2m} \left[ {\vec{\sigma}\cdot \left( {\vec{A}_{+}-\vec{A}_{-}}%
\right) }\right] \chi _1  \label{eq46}
\end{equation}
\begin{equation}
i\frac{\partial \chi _1}{\partial t}=\left[ {\frac 1{2m}\left[ {\left( {-i%
\vec{\nabla}-\vec{A}_{-}}\right) ^2+\delta ^2}\right] +A_{-,0}-\left( {\frac{%
\vec{\sigma}\cdot \vec{B}_{-}}{2m}}\right) }\right]  \chi
_1+i\frac
\delta {2m} \left[ {\vec{\sigma}\cdot \left( {\vec{A}_{-}-\vec{A}_{+}}%
\right) }\right]  \varphi _1  \label{eq47}
\end{equation}
with $\vec{B}_{+}$ and $\vec{B}_{-}$ the magnetic fields on the
two sheets, and where we have used $\psi _+ =\left( {{
\begin{array}{cc}
{\alpha _1 } &  \\
{\alpha _2 } &
\end{array}
}} \right)$ and $\psi _- =\left( {{
\begin{array}{cc}
{\beta _1 } &  \\
{\beta _2 } &
\end{array}
}} \right)$. To derive 33 and 34, we have assumed that the mass
term prevails on the kinetic and coulomb energies.

So, we get two Pauli equations coupled through the magnetic vector
potentials of both sheets. We stress that the wave functions on
the two sheets correspond to the ``large components'' of the Dirac
equation as in the standard approach.

Clearly, those equations allow a non gravitational interaction
between the two sectors. The nature of the coupling is however
different from the one discussed in earlier papers as we will see
hereafter [3]. Once again, we stress that for $A_+ =A_- =0$, the
coupling disappears and the two copies of space-time are
completely non-interacting (in flat geometry). In that case, the
field theory treatment reduces to the standard quantum mechanics
with two independent Pauli equations.

\section{Effect of a constant vector potential on a massive fermion}

To investigate the behavior of a particle belonging to this ``Z2
space-time'', we now look at a special but physically instructive
case of coupling between both sheets. We first rewrite the
equations (33) and (34) in a much simpler form:
\begin{equation}
i \partial _0\left( {{
\begin{array}{c}
{\varphi _1} \\
{\chi _1}
\end{array}
}}\right) =\left( {{
\begin{array}{cc}
{H_{+}} & W \\
{-W} & {H_{-}}
\end{array}
}}\right)  \left( {{
\begin{array}{c}
{\varphi _1} \\
{\chi _1}
\end{array}
}}\right) =H_{Z2}\left( {{
\begin{array}{c}
{\varphi _1} \\
{\chi _1}
\end{array}
}}\right)   \label{eq48}
\end{equation}
with $H_{Z2}$ the two sheeted Hamiltonian with components
\begin{equation}
H_{+}=\frac 1{2m}\left[ {\left(
{-i\vec{\nabla}-\vec{A}_{+}}\right)
^2+\delta ^2}\right] +A_{+,0}-\left( {\frac{\vec{\sigma}\cdot \vec{B}_{+}}{2m%
}}\right)   \label{eq49}
\end{equation}
\begin{equation}
H_{-}=\frac 1{2m}\left[ {\left(
{-i\vec{\nabla}-\vec{A}_{-}}\right)
^2+\delta ^2}\right] +A_{-,0}-\left( {\frac{\vec{\sigma}\cdot \vec{B}_{-}}{2m%
}}\right)   \label{eq50}
\end{equation}
\begin{equation}
W=i\frac \delta {2m} \left[ {\vec{\sigma}\cdot \left( {\vec{A}_{+}-\vec{%
A}_{-}}\right) }\right]   \label{eq51}
\end{equation}
To focus specifically on the coupling between both sheets, let us
consider the more restrictive Hamiltonian
\begin{equation}
H_{Z2}=\left( {{
\begin{array}{cc}
0 & W \\
{-W} & 0
\end{array}
}}\right)   \label{eq52}
\end{equation}
The calculation can be further simplified by taking zero scalar
potentials on both folds, just keeping the vector potential in the
$+$ sheet, i.e.
\begin{equation}
A_{+,z}=A_0,\text{ }A_{+,j\ne z}=0\text{\ and\ }A_{-,z}=0,\text{
}A_{-,j\ne z}=0  \label{eq53}
\end{equation}
In that case, the previous system of Pauli equations leads to
\begin{equation}
\partial _0\varphi _1=\frac \delta {2m}\sigma ^z {A_0}
 \chi _1  \label{eq54}
\end{equation}
\begin{equation}
\partial _0\chi _1=-\frac \delta {2m}\sigma ^z {A_0}
 \varphi _1  \label{eq55}
\end{equation}
Using $\varphi _1=\left( {{
\begin{array}{c}
{\varphi _1^{+}} \\
{\varphi _1^{-}}
\end{array}
}}\right) $ and $\chi _1=\left( {{
\begin{array}{c}
{\chi _1^{+}} \\
{\chi _1^{-}}
\end{array}
}}\right) $ with $\sigma ^z=\left( {{
\begin{array}{cc}
1 & 0 \\
0 & {-1}
\end{array}
}}\right) $, one gets
\begin{equation}
\frac{\partial \varphi _1^{+}}{\partial t}=\frac \delta {2m}A_0
\chi _1^{+},\text{ }\frac{\partial \varphi _1^{-}}{\partial
t}=-\frac \delta
{2m}A_0 \chi _1^{-},\text{ }\frac{\partial \chi _1^{+}}{\partial t}%
=-\frac \delta {2m}A_0 \varphi _1^{+},\text{ }\frac{\partial \chi _1^{-}%
}{\partial t}=+\frac \delta {2m}A_0 \varphi _1^{-}  \label{eq56}
\end{equation}
The general solution can be readily derived
\[
\varphi _1^{+}(t)=C_1 \sin \left( {\omega t}\right) +C_2 \cos
\left( {\omega t}\right)
\]
\begin{equation}
\varphi _1^{-}(t)=C_4 \cos \left( {\omega t}\right) +C_3 \sin
\left( {\omega t}\right)   \label{eq57}
\end{equation}
\[
\chi _1^{+}(t)=C_1 \cos \left( {\omega t}\right) -C_2 \sin \left( {%
\omega t}\right)
\]
\[
\chi _1^{-}(t)=-C_3 \cos \left( {\omega t}\right) +C_4 \sin \left(
{\omega t}\right)
\]
where we have set $\omega =\frac{\delta  A_0}{2m}$.

Assume that at $t=0$, we have $\varphi _1^{+}(0)=1$, $\varphi _1^{-}(0)=0$, $%
\chi _1^{+}(0)=0$, $\chi _1^{-}(0)=0$ (the particle is located in
the ``$+$'' sheet with a spin up) then we get
\begin{equation}
\varphi _1^{+}(t)=\cos \left( {\omega t}\right) \text{\ and\ }\chi
_1^{+}(t)=-\sin \left( {\omega t}\right)   \label{eq58}
\end{equation}
whereas the other components of the wave function vanish.

Hence, the probabilities of finding the particle in the ``$+$'' or
``$-$'' sheet are respectively
\begin{equation}
P_{+}=\left| {\varphi _1^{+}(t)}\right| ^2=\cos ^2\left( {\omega
t}\right)
\text{\ and\ }P_{-}=\left| {\chi _1^{+}(t)}\right| ^2=\sin ^2\left( {\omega t%
}\right)   \label{eq59}
\end{equation}
Therefore, we find that the particle oscillates between the two
sectors with a time periodicity depending on the intensity of the
magnetic vector potential. This result is the most striking one of
this paper as it suggests a possible exchange of matter between
the two sheets. It is however a very simplified model that does
not take into account the perturbative effects exerted by the
environment.

\section{Combination of a constant vector potential and a magnetic field}

We can get some insight onto the question of the environmental
effects by considering the case of a static magnetic field
$B_{+}=B_0\vec{e}_z$ superimposed to the vector potential (40). In
this approach, no magnetic field is postulated in the ``$-$``
sheet. Moreover, one considers that the
vector potential due to the magnetic field $B_{+}$ is much smaller than $%
A_{+}$ such that its contribution to the overall vector potential
can be completely neglected. Under those assumptions, it can be
easily shown that the equations (33) and (34) lead to
\begin{equation}
i\partial _0\varphi _1=-\frac{\sigma ^z B_0}{2m}+\frac{i\delta }{2m}%
\sigma ^z A_0 \chi _1  \label{eq60}
\end{equation}
\begin{equation}
i\partial _0\chi _1=-\frac{i\delta }{2m}\sigma ^z A_0 \varphi _1
\label{eq61}
\end{equation}
A procedure similar to the one developed in the preceding
paragraph allows then to find the solution
\begin{equation}
\varphi _1^{+}(t)=\exp \left[ {i\frac K2t}\right]  \left\{ {\cos
\left[ {\frac 12\sqrt{K^2+4\omega ^2}t}\right] +\frac
K{\sqrt{K^2+4\omega ^2}}\sin \left[ {\frac 12\sqrt{K^2+4\omega
^2}t}\right] }\right\}   \label{eq62}
\end{equation}
\begin{equation}
\chi _1^{+}(t)=-\exp \left[ {i\frac K2t}\right]  \frac{2\omega }{\sqrt{%
K^2+4\omega ^2}} \sin \left[ {\frac 12\sqrt{K^2+4\omega
^2}t}\right] \label{eq63}
\end{equation}
with $K=B_0/2m$. The other components of the field vanish. To
derive eq.49 and eq.50, we have used the initial conditions of the
third paragraph at time $t=0$.

Thus, in presence of a magnetic field, the probabilities $P_{+}$,
$P_{-}$ become
\begin{equation}
P_{+}=\left| {\varphi _1^{+}(t)}\right| ^2=1-\frac{4\omega ^2}{4\omega ^2+K^2%
}\sin ^2\left[ {\frac 12\sqrt{K^2+4\omega ^2}t}\right] ,\text{
}P_{-}=1-P_{+} \label{eq64}
\end{equation}
So, it appears that the period and the amplitude of the
oscillations decrease with the magnetic field. As a consequence,
the particle is confined to the sheet with strongly suppressed
oscillations. One can easily convince oneself that a similar
consideration holds when the magnetic field is substituted by a
scalar potential.

\section{Discussion}

Several models in brane theories, predict that massive particles
are able to leave the brane and propagate freely in the 5D bulk.
However, it is usually assumed that only highly energetic
particles can travel that way. Contrarily, in our approach, a low
energy particle can move in the 5D bulk as well by doing
oscillations between both spacetime sheets. Still should we
explain how locality and energy conservation could be satisfied in
such circumstances. Indeed, from the point of view of a
``one-sheeted'' observer, as we are, the behavior of such a
particle would be in conflict with every known physical
principles, the most noticeable ones being locality and energy
conservation. However, when rescaled at a ``two-sheeted'' level,
the problem disappears since the sum of the energies of both
sheets remains then constant during the oscillations. So, for an
hypothetical observer, able to see both sheets simultaneously, the
particle never disappears from the 5D bulk and the apparent energy
violation problem in 4D is only an artifact of low dimensionality.

In our model, the oscillation frequency depends on the particle
mass. Therefore, one may wonder what would happen for an ensemble
of particles embedded in a region of high vector potential. If all
interactions can be neglected, our model predicts that each
particle will undergo oscillations at a specific frequency
depending on its mass. The lightest particles will oscillate first
followed by the heavier particles. In the case of strongly
interacting particles however, the situation is completely
different. The presence of the other neighboring particles must be
taken into account to describe adequately what happens. A complete
treatment of this problem is of course far beyond the scope of
this paper. Nevertheless, the results obtained by considering the
effect of a magnetic field are a first step to mimic the
environmental effects. Our results suggest that the oscillations
are strongly suppressed in presence of an applied field similar to
the one that could be generated by neighboring particles. In fact,
each collision or exchange of particles results likely in a
damping of the amplitude of the oscillations as in the familiar
quantum Zeno effect. In such circumstances, the particle remains
perfectly localized in its sheet with a frozen oscillatory
behavior.

Despite the complications due to environmental effects, it seems
possible to design an experimental set-up to evidence the
oscillatory behavior. A first critical condition to be satisfied
is to limit the perturbations due to environment. That implies to
avoid the presence of any electromagnetic fields and to study
isolated particles only. Very intense vector potentials must also
be used to enhance the oscillation frequency between the two
sheets. All these very specific conditions can be met, for
example, in a hollow cylinder with an inner flow of electrical
current, operating under a high vacuum. In this system, no
magnetic nor electric field will be generated but nevertheless a
constant vector potential will appear in the hollow part of the
cylinder. A source emitting charged particles of low energy can
then be placed at one side of the system; ideally this source
should emit particles one by one to prevent their mutual
interaction. At the opposite side of the cylinder, a detector
should be placed to collect every particle emitted by the source.
If the vector potential is fixed at a sufficiently high value,
particles oscillations should occur and the detector should record
a lack of events.

If our model has a physical reality, various reasons can be
proposed to explain why ``two-sheeted oscillations'' have not been
observed so far

\begin{enumerate}
\item The damping of oscillations by collisional processes is so
huge that the fermions are completely trapped in their own
space-time sheet, \item The intensity of the usual electromagnetic
vector potentials are much too low to lead to observable
oscillations, \item The value of $\delta $ is so small that even
for very intense potentials, the oscillations remain very
difficult to be detected. Notice that since the distance between
the two sheets d varies like $1/\delta $, this would suggest that
both space-time-sheets are separated by a cosmological distance in
the discrete 5D bulk.
\end{enumerate}

Obviously, the last hypothesis is the most stimulating one.
Indeed, a cosmological model could shed a new light on several
puzzling phenomena such
as the missing mass matter, the acceleration of the cosmic expansion{\ldots}%
and could also provide an assessment of the distance between the
two sheets, necessary for determining the typical values of
$\omega $. However, the study of such a cosmological model is left
for future prospects.

\section{Conclusion}

In this paper, we have considered the quantum dynamics of massive
particles in a two-sheeted space-time using the non-commutative
geometry. The two-sheeted counterparts of the usual Dirac and
Pauli equations have been derived and the free field solutions of
the Z2-Dirac equation have been explicitly given. It is shown that
our approach provides a possible description of matter and hidden
sector matter, which are then localized on different sheets in the
discrete 5D bulk. The model predicts that each isolated massive
particle oscillates between the two space-time sheets in presence
of a constant magnetic vector potential.


\begin{thebibliography}{}

\bibitem{1}T. D. Lee and C. N. Yang, ``Question of Parity conservation in
weak interaction'', Phys. Rev. 104 (1956) 254

\bibitem{2}R. Foot, A.Y. Ignatiev {\&} R.R. Volkas, ``Physics of mirror
photons'', Phys.Lett. B503 (2001) 355-361

\bibitem{3}R. Foot, ``Neutrino oscillations and the exact parity model'',
Mod.Phys.Lett. A9 (1994) 169-180

\bibitem{4}R. Foot, ``Seven (and a half) reasons to believe in Mirror
Matter: From neutrino puzzles to the inferred Dark matter in the
Universe'', Acta Phys.Polon. B32 (2001) 2253-2270

\bibitem{5}M. Green, J. Schwarz and E. Witten, Superstring Theories,
Cambridge University press (1989).

\bibitem{6}A. Connes, Non-Commutative Geometry, [Academic Press, 1994].

\bibitem{7}A.Connes, J.Lott, Nucl.Phys. B18 (Proc.Suppl.) (1990) 29.

\bibitem{8}N.A.Viet, K.C.Wali, ``A Discretized Version of Kaluza-Klein
Theory with Torsion and Massive Fields'', Int. J. Modern Phys. A,
11(1996), 2403

\bibitem{9}N.A.Viet, K.C.Wali, ``Non-commutative geometry and a
Discretized Version of Kaluza-Klein theory with a finite field
content'', Int.J.Mod.Phys. A11 (1996) 533

\bibitem{10}J.M.Gracia-Bondia, B.Iochum {\&} T.Schucker, ``The standard
model in noncommutative geometry and fermion doubling'',
Phys.Lett. B416 (1998) 123-128

\bibitem{11}F.Lizzi, G.Mangano, G.Miele {\&} G.Sparano, ``Mirror Fermions
in Noncommutative Geometry'', Mod.Phys.Lett. A13 (1998) 231-23

\bibitem{12}N.A. Viet , K.C. Wali, ``Chiral spinors and gauge fields in
noncommutative curved space-time'', Phys.Rev. D67 (2003) 124029

\bibitem{13}E. Sassaroli, Flavor Oscillation in Field Theory,
hep-ph/9609476.

\bibitem{14}A. A. Prinz, R. Baggs, J. Ballam, S. Ecklund, C. Fertig, J. A.
Jaros, K. Kase, A. Kulikov, W. G. J. Langeveld, R. Leonard, T.
Marvin, T. Nakashima, W. R. Nelson, A. Odian, M. Pertsova, G.
Putallaz, and A. Weinstein, ``Search for Millicharged Particles at
SLAC", Phys. Rev. Lett. 81 (1998) 1175

\end{thebibliography}
\end{document}